\documentstyle[amstex,pra,aps]{revtex}
\twocolumn

\begin{document}
\draft
\title{A microscopic quantum dynamics approach to the dilute condensed Bose gas}
\author{Thorsten K\"ohler and Keith Burnett}
\address{Clarendon Laboratory, Department of Physics, \\
University of Oxford, Oxford, OX1 3PU, United Kingdom}
\date{\today}
\maketitle

\begin{abstract}
We derive quantum evolution equations for the dynamics of
dilute condensed Bose gases. The approach contains, at different 
orders of approximation, for cases close to equilibrium, 
the Gross Pitaevskii equation and the
first order Hartree Fock Bogoliubov theory. 
The proposed approach is also suited for the
description of the dynamics of condensed gases which are far away from
equilibrium. As an example the scattering of two Bose condensates is
discussed.
\end{abstract}


\section{\protect\bigskip Introduction}

The description of the evolution
of condensates far from equilibrium is now a matter of considerable
importance to the field of matter wave physics. This includes colliding 
\cite{Deng}
and
collapsing 
\cite{Roberts,Donley}
condensates where components of the overall condensed system
are moving rapidly with respect to one another. 
Numerous quantitative theoretical
studies of partially condensed gases to date have relied mostly on
the time dependent Gross-Pitaevskii approach and extended kinetic theories
(see, e.g., \cite{GZ5,HWC,WWCH})
that are appropriate to fluctuations and excitations around thermal
equilibrium situations. These general approaches all depend on the contact 
potential approximation of the binary interactions which restricts  
two body collisions to zero momenta. New methods are, therefore, required 
to describe systems outside this restricted domain.  

In this paper we shall describe a technique that
we believe can provide a comprehensive approach to the description of
evolving condensates. We use the method of non-commutative cumulants 
\cite {Fricke}
to derive a set of coupled equations that should provide an accurate
description in regimes where previous approaches fail. We shall explain the
formal basis of the method before applying it to the partially condensed
gas. We shall then focus on the specific case of a colliding pair of
condensates to show the utility of the method. The general coupled equations
should also be useful for the description of four-wave mixing experiments, 
for collapsing condensates and other cases where bulk
relative motion of condensates is observed. In the next section we shall
define the Hamiltonian for the systems we wish to study and find the formal
equations of motion for the correlation functions. 

\section{Quantum evolution equations for correlation functions}

\noindent 

A many body system of identical boson atoms with a pair interaction $V(%
{\bf r})$ is described by the Hamiltonian 
\begin{align}
& H=\int d{\bf x}\psi ^{\dagger }({\bf x})H_{{\rm 1B}}({\bf x})\psi ({\bf x})
\nonumber \\
& +\frac{1}{2}\int d{\bf x}_{1}\int d{\bf x}_{2}\psi ^{\dagger }({\bf x}%
_{1})\psi ^{\dagger }({\bf x}_{2})V({\bf x}_{1}-{\bf x}_{2})\psi ({\bf x}%
_{2})\psi ({\bf x}_{1}).  \label{Hamiltonian}
\end{align}
Here $H_{{\rm 1B}}({\bf x})$ is the one body Hamiltonian containing the
kinetic energy $-\hbar ^{2}\Delta _{{\bf x}}/2m$ and the trapping potential $%
V_{{\rm trap}}({\bf x})$. The field operators satisfy the boson commutation
relations 
\begin{align}
\left[ \psi ({\bf x}_{1}),\psi ({\bf x}_{2})\right] & =0,  \nonumber \\
\left[ \psi ({\bf x}_{1}),\psi ^{\dagger }({\bf x}_{2})\right] & =\delta (%
{\bf x}_{1}-{\bf x}_{2}).  \label{commrel}
\end{align}

Throughout this article the state of the many body system at time $t$ is
described by a statistical operator $\rho (t)$ and the expectation value of
an arbitrary operator $A$ is denoted by 
\begin{equation}
\langle A\rangle _{t}={\rm Tr}(\rho (t)A).  \label{defexpvalue}
\end{equation}
The Schr\"{o}dinger equation then yields the dynamic equation 
\begin{equation}
i\hbar \frac{\partial }{\partial t}\langle A\rangle _{t}=\langle \lbrack
A,H]\rangle _{t}.  \label{dynamicEqA}
\end{equation}
In the following, expectation values in Eq.~(\ref{defexpvalue}) are denoted
as correlation functions if $A$ is a product of creation or annihilation
operators. Since every expectation value can be expanded in linear
combinations of normal ordered correlation functions their dynamics
determines all physical properties of the many body system. According to
Eq.~(\ref{dynamicEqA}) the system of dynamic equations for the normal
ordered correlation functions up to the order of $n$ reads 
\begin{align}
& i\hbar \frac{\partial }{\partial t}\langle \psi ({\bf x}_{1})\rangle
_{t}=\langle \lbrack \psi ({\bf x}_{1}),H]\rangle _{t},  \nonumber \\
& i\hbar \frac{\partial }{\partial t}\langle \psi ({\bf x}_{2})\psi ({\bf x}%
_{1})\rangle _{t}=\langle \lbrack \psi ({\bf x}_{2})\psi ({\bf x}%
_{1}),H]\rangle _{t},  \nonumber \\
& i\hbar \frac{\partial }{\partial t}\langle \psi ^{\dagger }({\bf x}%
_{2})\psi ({\bf x}_{1})\rangle _{t}=\langle \lbrack \psi ^{\dagger }({\bf x}%
_{2})\psi ({\bf x}_{1}),H]\rangle _{t},  \nonumber \\
& \vdots   \nonumber \\
& i\hbar \frac{\partial }{\partial t}\langle \psi ^{\dagger }({\bf x}%
_{n})\cdots \psi ({\bf x}_{1})\rangle _{t}=\langle \lbrack \psi ^{\dagger }(%
{\bf x}_{n})\cdots \psi ({\bf x}_{1}),H]\rangle _{t}.  \label{kinEqcorr}
\end{align}

The system Eq.~(\ref{kinEqcorr}) is incomplete since the commutator of a
product of $n$ field operators with $H$ contains products of $n+2$ field
operators. The main problem thus consists in truncating Eq.~(\ref{kinEqcorr}%
) appropriately. An approximate method to close the kinetic equations at an
arbitrary order has been proposed by Fricke \cite{Fricke} in order to
describe the dynamics of many body systems on short time scales. In the
latter approach Eq.~(\ref{kinEqcorr}) is transformed into an equivalent
system of differential equations for what are termed non-commutative
cumulants. This article contains the extension of Fricke's work to the
description of the dynamics of dilute Bose gases and their equilibrium
properties. 

\section{Dynamic equations for non commutative cumulants}
\noindent
Given a set of boson creation and annihilation operators $%
B_{1},B_{2},\ldots $ the cumulant expansion of their respective correlation
functions is defined recursively by
\cite{Fricke} 
\begin{align}
& \langle B_{1}\rangle =\langle B_{1}\rangle ^{c},  \nonumber \\
& \langle B_{1}B_{2}\rangle =\langle B_{1}B_{2}\rangle ^{c}+\langle
B_{1}\rangle ^{c}\langle B_{2}\rangle ^{c},  \nonumber \\
& \langle B_{1}B_{2}B_{3}\rangle =\langle B_{1}B_{2}B_{3}\rangle
^{c}+\langle B_{1}\rangle ^{c}\langle B_{2}B_{3}\rangle ^{c}  \nonumber \\
& \quad +\langle B_{2}\rangle ^{c}\langle B_{1}B_{3}\rangle ^{c}+\langle
B_{3}\rangle ^{c}\langle B_{1}B_{2}\rangle ^{c}+\langle B_{1}\rangle
^{c}\langle B_{2}\rangle ^{c}\langle B_{3}\rangle ^{c},  \nonumber \\
& \vdots   \label{defcumI}
\end{align}
The expressions $\langle B_{1}\cdot \ldots \cdot B_{n}\rangle ^{c}$ define
the non commutative cumulants we shall use. The cumulants can be obtained
directly in terms of correlation functions through 
\begin{align}
&\langle B_{1}\cdots B_{n}\rangle ^{c}
\nonumber\\
&=\frac{\partial }{\partial x_{1}}\cdots 
\frac{\partial }{\partial x_{n}}\log \left. \left\langle
e^{x_{1}B_{1}}\cdots e^{x_{n}B_{n}}\right\rangle 
\right| _{x_1=0,\ldots, x_n=0}.
\label{defcumII}
\end{align}

In contrast to their respective correlation functions the cumulants should
decrease with increasing order, as long as the system is not
too far away from the interaction free equilibrium. More precisely, 
for an ideal Bose gas in thermal equilibrium all cumulants containing more 
than two field operators vanish. In fact, if the state of an ideal gas
is given by the grand canonical density matrix, according to Wick's theorem 
of statistical mechanics \cite{FetterWalecka}, 
every number conserving normal ordered correlation function 
$\langle B_{1}\cdots B_{n}\rangle $ splits into a sum of all possible 
products of contractions $\langle B_{i}B_{j}\rangle $ which preserve the 
initial order of the operators. Since the expectation value of a single 
field operator vanishes, the contractions equal their respective second 
order cumulants. Inserting the Wick decompositions of the correlation 
functions into the left hand sides of Eqs.~(\ref{defcumI}) shows 
successively that only the number conserving cumulants of the second order 
can be different from zero.
In this sense the higher order cumulants are a measure of the deviations 
of the physical system from the interaction free equilibrium.

Among other general properties it is worth mentioning that in the absence of
a pair interaction the cumulants satisfy the same dynamic equations as their
respective correlation functions. In addition, all field operators in $%
\langle B_1\cdots B_n\rangle^c$ commute as soon as the order $n$ exceeds
two, i.e. 
\begin{equation}
\langle\cdots B_iB_{i+1} \cdots\rangle^c = \langle\cdots
B_{i+1}B_i\cdots\rangle^c.
\end{equation}
According to Eq.~(\ref{defcumII}) the cumulants of the order of $n>1$ depend
only on the ``deviation operators'' $B_i-\langle B_i\rangle$, i.e. 
\begin{equation}
\langle B_1\cdots B_n\rangle^c= \langle (B_1-\langle B_1\rangle)\cdots
(B_n-\langle B_n\rangle)\rangle^c.
\end{equation}
Hence, the cumulant expansion Eq.~(\ref{defcumI}) is also useful for the
description of Bose condensed systems with non number conserving correlation
functions.

Due to their particular roles in the description of condensed
dilute Bose gases throughout this article, the first and second order normal
ordered cumulants are denoted as follows: 
\begin{align}
\Psi ({\bf x},t)& =\langle \psi ({\bf x})\rangle _{t}^{c},  \nonumber \\
\Phi ({\bf x}_{1},{\bf x}_{2},t)& =\langle \psi ({\bf x}_{2})\psi ({\bf x}%
_{1})\rangle _{t}^{c},  \nonumber \\
\Gamma ({\bf x}_{1},{\bf x}_{2},t)& =\langle \psi ^{\dagger }({\bf x}%
_{2})\psi ({\bf x}_{1})\rangle _{t}^{c}.
\end{align}
The first order cumulant $\Psi ({\bf x},t)$ is usually referred to as the
condensate wave function and $\Phi ({\bf x}_{1},{\bf x}_{2},t)$ as the pair
function. $\Gamma ({\bf x}_{1},{\bf x}_{2},t)$ is interpreted as the one
body density matrix of the non condensed fraction. The total one body
correlation function then reads 
\begin{align}
\rho ^{(1)}({\bf x}_{1},{\bf x}_{2},t)& =\langle \psi ^{\dagger }({\bf x}%
_{2})\psi ({\bf x}_{1})\rangle _{t}  \nonumber \\
& =\Gamma ({\bf x}_{1},{\bf x}_{2},t)+\Psi ({\bf x}_{1},t)\Psi ^{\ast }({\bf %
x}_{2},t).
\end{align}
Hence, the density of the gas at the position ${\bf x}$ and time $t$ is
given by $\rho ^{(1)}({\bf x},{\bf x},t)$.

The dynamic equations for the normal ordered cumulants can be derived from
Eqs.~(\ref{kinEqcorr}) by commuting successively all creation operators in
the commutators on the right hand sides to the left and expanding the
correlation functions on both sides according to Eq.~(\ref{defcumI}). Hence,
the exact dynamics of the condensate wave function is described by 
\begin{align}
& i\hbar \frac{\partial }{\partial t}\Psi ({\bf x},t)=H_{{\rm 1B}}({\bf x}%
)\Psi ({\bf x},t)  \nonumber \\
& +\int d{\bf y}V({\bf x}-{\bf y})[\langle \psi ^{\dagger }({\bf y})\psi (%
{\bf y})\psi ({\bf x})\rangle _{t}^{c}+\Psi ({\bf y},t)\Gamma ({\bf x},{\bf y%
},t)  \nonumber \\
& \qquad +\Psi ({\bf x},t)\Gamma ({\bf y},{\bf y},t)]  \nonumber \\
& +\int d{\bf y}V({\bf x}-{\bf y})\Psi ^{\ast }({\bf y},t)\left[ \Phi ({\bf x%
},{\bf y},t)+\Psi ({\bf x},t)\Psi ({\bf y},t)\right] .  \label{exdyneqord}
\end{align}
In the same way the exact equations of motion for the pair function 
\begin{align}
& i\hbar \frac{\partial }{\partial t}\Phi ({\bf x}_{1},{\bf x}_{2},t)=H_{%
{\rm 2B}}({\bf x}_{1},{\bf x}_{2})\Phi ({\bf x}_{1},{\bf x}_{2},t)  \nonumber
\\
& +V({\bf x}_{1}-{\bf x}_{2})\Psi ({\bf x}_{1},t)\Psi ({\bf x}_{2},t) 
\nonumber \\
& +\bigg\{\int d{\bf y}V({\bf x}_{1}-{\bf y})[\langle \psi ^{\dagger }({\bf y%
})\psi ({\bf y})\psi ({\bf x}_{2})\psi ({\bf x}_{1})\rangle _{t}^{c} 
\nonumber \\
& \qquad +\Psi ({\bf y},t)\langle \psi ^{\dagger }({\bf y})\psi ({\bf x}%
_{2})\psi ({\bf x}_{1})\rangle _{t}^{c}  \nonumber \\
& \qquad +\Psi ({\bf x}_{1},t)\langle \psi ^{\dagger }({\bf y})\psi ({\bf y}%
)\psi ({\bf x}_{2})\rangle _{t}^{c}  \nonumber \\
& \qquad +\Gamma ({\bf y},{\bf y},t)\Phi ({\bf x}_{1},{\bf x}_{2},t)+\Gamma (%
{\bf x}_{1},{\bf y},t)\Phi ({\bf x}_{2},{\bf y},t)]  \nonumber \\
& \quad +\int d{\bf y}V({\bf x}_{1}-{\bf y})\Psi ^{\ast }({\bf y},t)[\langle
\psi ({\bf y})\psi ({\bf x}_{2})\psi ({\bf x}_{1})\rangle _{t}^{c}  \nonumber
\\
& \qquad +\Psi ({\bf y},t)\Phi ({\bf x}_{1},{\bf x}_{2},t)+\Psi ({\bf x}%
_{1},t)\Phi ({\bf x}_{2},{\bf y},t)]  \nonumber \\
& \quad +\int d{\bf y}V({\bf x}_{1}-{\bf y})\Gamma ({\bf x}_{2},{\bf y},t) 
\nonumber \\
& \qquad \times \lbrack \Phi ({\bf x}_{1},{\bf y},t)+\Psi ({\bf x}%
_{1},t)\Psi ({\bf y},t)]\bigg\}  \nonumber \\
& +\{{\bf x}_{1}\leftrightarrow {\bf x}_{2}\}  \label{exdyneqanom}
\end{align}
and the density matrix of the non condensed fraction 
\begin{align}
& i\hbar \frac{\partial }{\partial t}\Gamma ({\bf x}_{1},{\bf x}_{2},t)=[H_{%
{\rm 1B}}({\bf x}_{1})-H_{{\rm 1B}}({\bf x}_{2})]\Gamma ({\bf x}_{1},{\bf x}%
_{2},t)  \nonumber \\
& +\bigg\{\int d{\bf y}V({\bf x}_{1}-{\bf y})[\langle \psi ^{\dagger }({\bf x%
}_{2})\psi ^{\dagger }({\bf y})\psi ({\bf y})\psi ({\bf x}_{1})\rangle
_{t}^{c}  \nonumber \\
& \qquad +\Psi ({\bf y},t)\langle \psi ^{\dagger }({\bf x}_{2})\psi
^{\dagger }({\bf y})\psi ({\bf x}_{1})\rangle _{t}^{c}  \nonumber \\
& \qquad +\Psi ({\bf x}_{1},t)\langle \psi ^{\dagger }({\bf x}_{2})\psi
^{\dagger }({\bf y})\psi ({\bf y})\rangle _{t}^{c}  \nonumber \\
& \qquad +\Gamma ({\bf y},{\bf x}_{2},t)\Gamma ({\bf x}_{1},{\bf y}%
,t)+\Gamma ({\bf x}_{1},{\bf x}_{2},t)\Gamma ({\bf y},{\bf y},t)]  \nonumber
\\
& \quad +\int d{\bf y}V({\bf x}_{1}-{\bf y})\Psi ^{\ast }({\bf y},t)[\langle
\psi ^{\dagger }({\bf x}_{2})\psi ({\bf y})\psi ({\bf x}_{1})\rangle _{t}^{c}
\nonumber \\
& \qquad +\Psi ({\bf x}_{1},t)\Gamma ({\bf y},{\bf x}_{2},t)+\Psi ({\bf y}%
,t)\Gamma ({\bf x}_{1},{\bf x}_{2},t)]  \nonumber \\
& \quad +\int d{\bf y}V({\bf x}_{1}-{\bf y})\Phi ^{\ast }({\bf y},{\bf x}%
_{2},t)  \nonumber \\
& \qquad \times \lbrack \Phi ({\bf x}_{1},{\bf y},t)+\Psi ({\bf x}%
_{1},t)\Psi ({\bf y},t)]\bigg\}  \nonumber \\
& -\{{\bf x}_{1}\leftrightarrow {\bf x}_{2}\}^{\ast }  \label{exdyneqnorm}
\end{align}
are obtained. Here $\{{\bf x}_{1}\leftrightarrow {\bf x}_{2}\}$ denotes the
exchange of the coordinates ${\bf x}_{1}$ and ${\bf x}_{2}$ in the bracket
and 
\begin{equation}
H_{{\rm 2B}}({\bf x}_{1},{\bf x}_{2})=H_{{\rm 1B}}({\bf x}_{1})+H_{{\rm 1B}}(%
{\bf x}_{2})+V({\bf x}_{1}-{\bf x}_{2})
\end{equation}
denotes the two body Hamiltonian. 

As proposed by Fricke \cite{Fricke}, a
first approach to close the dynamic equations for cumulants could consist in
keeping all cumulants up to a certain order $n$. If, for example, $n=1$ is
chosen Eq.~(\ref{exdyneqord}) yields the Gross Pitaevskii equation in the
Born approximation, 
\begin{align}
i\hbar \frac{\partial }{\partial t}\Psi ({\bf x},t)=& \left[ H_{{\rm 1B}}(%
{\bf x})+\int d{\bf y}V({\bf x}-{\bf y})|\Psi ({\bf y},t)|^{2}\right]  
\nonumber \\
& \times \Psi ({\bf x},t).  \label{GPEBorn}
\end{align}
Neglecting all cumulants containing three or four field operators in Eqs.~(%
\ref{exdyneqord}), (\ref{exdyneqanom}) and (\ref{exdyneqnorm}) yields a
system of time dependent first order Hartree Fock Bogoliubov equations 
which constitute the second order approximation. 
An alternative derivation of these second order equations, 
in the contact potential approximation, which includes applications 
to the Timmermans model of molecular formation in condensates 
\cite{Timmermans}
can be found 
in Ref.~\cite{HollandPRL}.
An independent approach which contains also the
cumulants up to the third order has been proposed by Proukakis and 
one of the authors 
\cite{ProukakisBurnett}.

These methods neglect multiple scattering in the equations for the last two
relevant cumulants and can be interpreted as an expansion in terms of $%
V(t-t_{0})/\hbar$, where $t-t_{0}$ is the time passed since the initial
time. For this reason they are applicable mainly for the description of
short time dynamics \cite{Fricke}. The multiple scattering contributions are
contained in the dynamic equations for the cumulants of the order of $n+1$, $%
n+2$, etc.. 

A systematic extension of the previous methods, which accounts
for multiple scattering in all equations of motion for the normal ordered
cumulants up to the order of $n$, can be achieved by including the free
dynamics of the normal ordered cumulants of the order of $n+1$ and $n+2$.
Here, free dynamics means neglecting all products of normal ordered
cumulants containing $n+3$ or $n+4$ field operators in the equations of
motion for the normal ordered cumulants of the order of $n+1$ and $n+2$. As
will be shown in Subsection \ref{subsec:equilibrium}, including multiple
scattering in all equations of motion at a given order of approximation
significantly extends their range of validity.

\section{First order approach}

\label{FirstOrder} \noindent In this section the extended cumulant method
will be illustrated for the lowest order $n=1$ \ \ \ approximation. As in
the case of the related Gross Pitaevskii approach, the resulting lowest
order dynamic equation is applicable only as long as the influence of the
non condensed fraction on the condensate can be neglected. This, in turn,
restricts the initial states of the gas we can treat with this approximation.

Neglecting all products of cumulants containing four or five field operators
in the equations of motion for the cumulants of the order of two and three,
the dynamic equation for the condensate wave function is given by Eq.~(\ref
{exdyneqord}). According to Eqs.~(\ref{exdyneqanom}) and (\ref{exdyneqnorm})
the respective equations for $\Phi $ and $\Gamma $ in the same approximation
become 
\begin{align}
i\hbar \frac{\partial }{\partial t}\Phi ({\bf x}_{1},{\bf x}_{2},t)=& H_{%
{\rm 2B}}({\bf x}_{1},{\bf x}_{2})\Phi ({\bf x}_{1},{\bf x}_{2},t)  \nonumber
\\
& +V({\bf x}_{1}-{\bf x}_{2})\Psi ({\bf x}_{1},t)\Psi ({\bf x}_{2},t)
\label{appr2,0}
\end{align}
and 
\begin{equation}
i\hbar \frac{\partial }{\partial t}\Gamma ({\bf x}_{1},{\bf x}_{2},t)=[H_{%
{\rm 1B}}({\bf x}_{1})-H_{{\rm 1B}}({\bf x}_{2})]\Gamma ({\bf x}_{1},{\bf x}%
_{2},t).  \label{appr1,1}
\end{equation}
Among all third order normal ordered cumulants, only those which appear on
the right hand side of Eq.~(\ref{exdyneqord}) have to be considered. We
first expand the dynamic equation for the correlation function $\langle \psi
^{\dagger }({\bf x}_{3})\psi ({\bf x}_{2})\psi ({\bf x}_{1})\rangle _{t}$
according to Eq.~(\ref{defcumI}) and neglect products of normal ordered
cumulants containing five field operators. The equation of motion for the
only relevant normal ordered\ third order cumulant is then given by: 
\begin{align}
& i\hbar \frac{\partial }{\partial t}\langle \psi ^{\dagger }({\bf x}%
_{3})\psi ({\bf x}_{2})\psi ({\bf x}_{1})\rangle _{t}^{c}  \nonumber \\
& =[H_{{\rm 2B}}({\bf x}_{1},{\bf x}_{2})-H_{{\rm 1B}}({\bf x}_{3})]\langle
\psi ^{\dagger }({\bf x}_{3})\psi ({\bf x}_{2})\psi ({\bf x}_{1})\rangle
_{t}^{c}  \nonumber \\
& \quad +V({\bf x}_{1}-{\bf x}_{2})\big[\Psi ({\bf x}_{2},t)\Gamma ({\bf x}%
_{1},{\bf x}_{3},t)  \nonumber \\
& \qquad \qquad \qquad \quad +\Psi ({\bf x}_{1},t)\Gamma ({\bf x}_{2},{\bf x}%
_{3},t)\big].  \label{appr2,1}
\end{align}

Equations (\ref{appr2,0}), (\ref{appr1,1}) and (\ref{appr2,1}) can be solved
formally in terms of the time dependent two body Green's function. A
detailed description of the Green's function methods applied in this article
is given in Ref.~\cite{Newton}. The retarded two body Green's function is
related to the respective time development operator $U_{{\rm 2B}%
}(t,t_{0})=\exp [-iH_{{\rm 2B}}(t-t_{0})/\hbar ]$ through 
\begin{equation}
G_{{\rm 2B}}^{(+)}(t,t_{0})=\frac{1}{i\hbar }\theta (t-t_{0})U_{{\rm 2B}%
}(t,t_{0}),  \label{defGreen'sfuncU}
\end{equation}
where $\theta (t-t_{0})$ is the step function which yields unity for $t>t_{0}
$ and vanishes elsewhere. Equation (\ref{defGreen'sfuncU}) shows that $G_{%
{\rm 2B}}^{(+)}$ obeys the equation: 
\begin{equation}
\left( i\hbar \frac{\partial }{\partial t}-H_{{\rm 2B}}\right) G_{{\rm 2B}%
}^{(+)}(t,t_{0})=\delta (t-t_{0}).  \label{defGreen'sfunc}
\end{equation}
A convenient representation of the formal solutions to Eqs.~(\ref{appr2,0}),
(\ref{appr1,1}) and (\ref{appr2,1}) is achieved by changing from position
space to the basis defined by the trap states $\phi _{i}$. 
In this representation the field
operators $\psi $ become single mode annihilation operators $a_{i}=\int d%
{\bf x}\phi _{i}^{\ast }({\bf x})\psi ({\bf x}),$ and the relevant cumulants
are given by $\Psi _{i}(t)=\langle a_{i}\rangle _{t}^{c}$, $\Phi
_{ij}(t)=\langle a_{j}a_{i}\rangle _{t}^{c}$ and $\Gamma _{ij}(t)=\langle
a_{j}^{\dagger }a_{i}\rangle _{t}^{c}$. Using Eq.~(\ref{defGreen'sfunc}) the
formal solution for the pair function in Eq.~(\ref{appr2,0}) can be written
in the form: 
\begin{align}
& \Phi _{ij}(t)=\sum_{k_{1}k_{2}}{_{s}}\langle i,j|U_{{\rm 2B}%
}(t,t_{0})|k_{1},k_{2}\rangle _{s}\Phi _{k_{1}k_{2}}(t_{0})  \nonumber \\
& +\sum_{k_{1}k_{2}}\int_{t_{0}}^{t}dt_{1}{_{s}}\langle i,j|G_{{\rm 2B}%
}^{(+)}(t,t_{1})V|k_{1},k_{2}\rangle _{s}\Psi _{k_{1}}(t_{1})\Psi
_{k_{2}}(t_{1}).  \label{apprcum2,0}
\end{align}
Equation (\ref{appr1,1}) yields the density matrix 
of the non condensed fraction thus: 
\begin{equation}
\Gamma _{ij}(t)=e^{-i(E_{i}-E_{j})(t-t_{0})/\hbar }\Gamma _{ij}(t_{0}).
\label{apprcum1,1}
\end{equation}
In the same way, the formal solution for the third order cumulant in Eq.~(%
\ref{appr2,1}) becomes 
\begin{align}
\langle a_{k}^{\dagger }a_{j}a_{i}\rangle _{t}^{c}=& \sum_{k_{1}k_{2}}{_{s}}%
\langle i,j|U_{{\rm 2B}}(t,t_{0})|k_{1},k_{2}\rangle _{s}  \nonumber \\
& \quad \times \langle a_{k}^{\dagger }a_{k_{2}}a_{k_{1}}\rangle
_{t_{0}}^{c}e^{iE_{k}(t-t_{0})/\hbar }  \nonumber \\
& +\sum_{k_{1}k_{2}}\int_{t_{0}}^{t}dt_{1}{_{s}}\langle i,j|G_{{\rm 2B}%
}^{(+)}(t,t_{1})V|k_{1},k_{2}\rangle _{s}  \nonumber \\
& \quad \times \Psi _{k_{1}}(t_{1})\Gamma
_{k_{2}k}(t_{1})e^{iE_{k}(t-t_{1})/\hbar }.  \label{apprcum2,1}
\end{align}
Here $t_{0}$ denotes the initial time and $|i,j\rangle _{s}$ is the
(not necessarily normalized)
symmetrized product of trap states $[\phi _{i}({\bf x}_{1})\phi _{j}({\bf x}%
_{2})+\phi _{i}({\bf x}_{2})\phi _{j}({\bf x}_{1})]/2$. According to Eqs.~(%
\ref{apprcum1,1}) and (\ref{apprcum2,1}) in the first order cumulant
approach, the density matrix of the non condensed fraction $\Gamma _{ij}(t)$
and the third order cumulant $\langle a_{k}^{\dagger }a_{j}a_{i}\rangle
_{t}^{c}$ will not evolve in time as long as they vanish initially. Their
contributions can therefore be neglected.

Equations (\ref{apprcum2,0}), (\ref{apprcum1,1}) and (\ref{apprcum2,1})
inserted into Eq.~(\ref{exdyneqord}) yield the closed nonlinear Schr\"{o}%
dinger equation for the condensate wave function 
\begin{align}
& i\hbar \frac{\partial }{\partial t}\Psi _{i}(t)=E_{i}\Psi _{i}(t) 
\nonumber \\
& \quad +\sum_{k_{1}k_{2}k_{3}}\int_{t_{0}}^{\infty }dt_{1}{_{s}}\langle
i,k_{3}|T_{{\rm 2B}}^{(+)}(t,t_{1})|k_{1},k_{2}\rangle _{s} 
\nonumber\\
& \qquad \times \Psi _{k_{1}}(t_{1})\Psi _{k_{2}}(t_{1})\Psi _{k_{3}}^{\ast
}(t).  \label{NLSnonMarkovEnergBasis}
\end{align}
Here, $T_{{\rm 2B}}^{(+)}$ denotes the time dependent and retarded two body
transition matrix 
\begin{equation}
T_{{\rm 2B}}^{(+)}(t,t_{0})=V\delta (t-t_{0})+VG_{{\rm 2B}}^{(+)}(t,t_{0})V.
\label{timedepTmatrix}
\end{equation}
The initial pair function $\Phi _{k_{1}k_{2}}(t_{0})$, which accounts for
collisions between condensate atoms that occur before $t_{0}$, is neglected
in Eq.~(\ref{NLSnonMarkovEnergBasis}). If the condensate is prepared as an
ideal gas ground state $\Phi _{k_{1}k_{2}}(t_{0})$ vanishes exactly \cite
{coherentstate}. Recent experiments using magnetic fields to manipulate the
pair interaction \cite{Cornish} \ have shown how to prepare such an initial
state of a Bose condensed gas. 
In general, the state of the gas will include these initial correlations
produced by collisions before the initial time we are considering.

It is worth noting that according to Eqs.~(\ref{timedepTmatrix}) and (\ref
{defGreen'sfuncU}) in the time integral on the right hand side of Eq.~(\ref
{NLSnonMarkovEnergBasis}) the condensate wave function is evaluated only at
the present time $t_{1}=t$ or in the past. This so called non-Markovian
property of the nonlinear Schr\"{o}dinger equation results from the
influence of the higher order cumulants on the condensate, which modifies 
its dynamics.

\subsection{Equilibrium properties}

\label{subsec:equilibrium} \noindent To describe many present day
experiments we can use the Markov
%({\em I am not sure about the
%spelling of Markoff}. {\bf The spelling of this name is certainly not clear
%since it is a result of translation from Russian language. I think that 
%``Markov'' is the most common version (see also
%http://www-groups.dcs.st-andrews.ac.uk/~history/Mathematicians/Markov.html))}
approximation in Eq.~(\ref{NLSnonMarkovEnergBasis}). In order
to obtain the Markov limit, the condensate is assumed to be in an ideal gas
initial state and the condensate wave function is transformed to the
interaction picture 
\begin{equation}
\Psi _{i}(t)=\Psi _{i}^{{\rm I}}(t)e^{-iE_{i}t/\hbar }.
\end{equation}
For a dilute gas the non Markovian interaction term on the right hand side
of Eq.~(\ref{NLSnonMarkovEnergBasis}) can be considered as a small
perturbation of the first, i.e.~ideal gas, contribution. This in turn leads
to a weak time dependence of $\Psi _{i}^{{\rm I}}(t)$. In the interaction
picture representation Eq.~(\ref{NLSnonMarkovEnergBasis}) reads 
\begin{align}
& i\hbar \frac{\partial }{\partial t}\Psi _{i}^{{\rm I}}(t)=%
\sum_{k_{1}k_{2}k_{3}}\int_{t_{0}}^{\infty }dt_{1}{_{s}}\langle i,k_{3}|T_{%
{\rm 2B}}^{(+)}(t,t_{1})|k_{1},k_{2}\rangle _{s}  \nonumber \\
& \times e^{-i(E_{k_{1}}+E_{k_{2}})t_{1}/\hbar }\Psi _{k_{1}}^{{\rm I}%
}(t_{1})\Psi _{k_{2}}^{{\rm I}}(t_{1})[\Psi _{k_{3}}^{{\rm I}}(t)]^{\ast
}e^{i(E_{k_{3}}+E_{i})t/\hbar }.  \label{IntPicNLS}
\end{align}
According to Eq.~(\ref{timedepTmatrix}) the time dependent transition matrix 
$T_{{\rm 2B}}^{(+)}(t,t_{1})$ is sharply peaked at $t_{1}=t$ with a width
determined by the two body collisional duration. As $\Psi _{i}^{{\rm I}}(t)$
is slowly varying on this time scale, all interaction picture condensate
wave functions can be evaluated at $t_{1}=t$ in Eq.~(\ref{IntPicNLS}). The
remaining time integral in Eq.~(\ref{IntPicNLS}) then simply contains a
Fourier transform of the time dependent transition matrix Eq.~(\ref
{timedepTmatrix}) and plays the role of a coupling function 
\begin{align}
{\cal T}_{{\rm 2B}}(i,k_{3};k_{1},k_{2})=& \int_{-\infty }^{t-t_{0}}d\tau {%
_{s}}\langle i,k_{3}|T_{{\rm 2B}}^{(+)}(t,t-\tau )|k_{1},k_{2}\rangle _{s} 
\nonumber \\
& \quad \times e^{i(E_{k_{1}}+E_{k_{2}})\tau /\hbar }.
\label{defcoupfunctionNLS}
\end{align}
Here, ${\cal T}_{{\rm 2B}}$ becomes approximately independent of time as
soon as $t-t_{0}$ exceeds the two body collisional duration. 
According to Eq.~(\ref{timedepTmatrix}) the time dependent transition 
matrix $T_{{\rm 2B}}^{(+)}$
is related to the usual two body $T$ matrix \cite{Newton} through 
\begin{equation}
T_{{\rm 2B}}^{(+)}(t,t-\tau )=\frac{1}{2\pi \hbar }\int dEe^{-iE\tau /\hbar
}T_{{\rm 2B}}(E+i0),  \label{defTmatrix}
\end{equation}
which still contains the trapping potential. Here ``$i0$'' denotes an
imaginary energy ``$i\varepsilon $'', where the positive parameter $%
\varepsilon $ is taken to zero after performing the energy integral. In many
present day experiments the trapping potential is nearly constant on the
scale of the spatial range of the two body interaction potential, so the two
body $T$ matrix in Eq.~(\ref{defTmatrix}) can be replaced by its free space
counterpart $T_{{\rm 2B}}^{{\rm free}}$. The time limit $t-t_{0}\rightarrow
\infty $ in Eq.~(\ref{defcoupfunctionNLS}) then yields 
\begin{equation}
{\cal T}_{{\rm 2B}}(i,k_{3};k_{1},k_{2})={_{s}}\langle i,k_{3}|T_{{\rm 2B}}^{%
{\rm free}}(E_{k_{1}}+E_{k_{2}}+i0)|k_{1},k_{2}\rangle _{s}.
\label{MarkovCoupfunctionNLS}
\end{equation}
As a consequence, the condensate wave function $\Psi _{i}(t)$ Eq.~(\ref
{IntPicNLS}) obeys the following form of the Gross Pitaevskii equation: 
\begin{align}
& i\hbar \frac{\partial }{\partial t}\Psi _{i}(t)=E_{i}\Psi _{i}(t) 
\nonumber \\
& \quad +\sum_{k_{1}k_{2}k_{3}}{\cal T}_{{\rm 2B}}(i,k_{3};k_{1},k_{2})\Psi
_{k_{1}}(t)\Psi _{k_{2}}(t)\Psi _{k_{3}}^{\ast }(t).  \label{MarkovNLS}
\end{align}
The coupling function in Eq.~(\ref{MarkovCoupfunctionNLS}), however,
exhibits an imaginary part which describes collisional losses of
energetically excited condensate atoms. The physical significance of this
imaginary part will be explained in Subsection \ref{subsec:Scatt}.

Close to equilibrium, the energy of occupied modes can be considered to be
so small that the coupling function is purely real to a good approximation.
The two body $T$ matrix can then be evaluated by means of the
contact potential approximation 
\begin{align}
T_{{\rm 2B}}^{{\rm free}}=& \int d{\bf p}_{1}d{\bf p}_{2}d{\bf p}_{3}d{\bf p}%
_{4}\delta ({\bf p}_{4}+{\bf p}_{3}-{\bf p}_{1}-{\bf p}_{2})  \nonumber \\
& \times |{\bf p}_{4}\rangle |{\bf p}_{3}\rangle \frac{4\pi \hbar ^{2}}{m}%
a_{0}\langle {\bf p}_{1}|\langle {\bf p}_{2}|.  \label{zerorangeappr}
\end{align}
Here $|{\bf p}_{i}\rangle $ denotes a one body plane wave momentum state and 
$m$ the atomic mass while $a_{0}$ is the two body $s$ wave scattering
length. In the contact potential approximation
Eq.~(\ref{MarkovNLS}) can be shown
to be formally equivalent to the Gross Pitaevskii equation in the Born
approximation (see Eq.~(\ref{GPEBorn})) with the two body potential $V({\bf r%
})$ replaced by $4\pi \hbar ^{2}a_{0}\delta ({\bf r})/m$. Hence, in the
position representation, the equilibrium limit of the Markovian nonlinear
Schr\"{o}dinger equation in Eq.~(\ref{MarkovNLS}) has the form:
\begin{equation}
i\hbar \frac{\partial }{\partial t}\Psi ({\bf x},t)=\left[ H_{{\rm 1B}}({\bf %
x})+\frac{4\pi \hbar ^{2}}{m}a_{0}|\Psi ({\bf x},t)|^{2}\right] \Psi ({\bf x}%
,t).  \label{GPE}
\end{equation}
This Gross Pitaevskii equation is transformed to its time independent form
by means of the ansatz 
$\Psi ({\bf x},t)=\Psi ({\bf x})\exp (-i\mu t/\hbar )$, 
where the energy parameter $\mu $ is usually termed the chemical
potential. 
The Markov approximation of the first order cumulant approach thus leads to
the correct Gross Pitaevskii equation, which describes virtually all
equilibrium phenomena in present day experiments with Bose condensates near
zero temperature.  

\subsection{Scattering of two condensates}

\label{subsec:Scatt} \noindent In this subsection, the cumulant approach
will be applied to describe the scattering of two condensates: a clear and
important non-equilibrium case. In a typical experiment the different
condensates are generated from a trapped parent condensate, at nearly zero
temperature, by means of an optical standing wave pulse. The trap in some
experiments can be switched off directly after the application of the light
pulse at time $t_{0}$. The photon momentum ${\bf p}_{{\rm ph}}$ can be
transferred to the atoms and  
is assumed to be much larger than the momentum
spread in the parent condensate and also the momentum $mc$ associated with
the speed of sound $c$. This implies that the condensate
wave function prepared in this
way contains three well separated momentum components with the respective
central 
atomic momenta ${\bf p}=\pm 2{\bf p}_{{\rm ph}}$ and ${\bf p}=0$. 
For simplicity, the
following analysis is restricted to the presence of two distinct components.
The condensate wave function can then be represented in the center of mass
frame as \cite{Band} 
\begin{align}
\Psi ({\bf x},t_{0})=& \left[ \Psi _{+}({\bf x},t_{0})e^{i{\bf p}_{{\rm ph}%
}\cdot {\bf x}/\hbar }+\Psi _{-}({\bf x},t_{0})e^{-i{\bf p}_{{\rm ph}}\cdot 
{\bf x}/\hbar }\right]   \nonumber \\
& \times e^{-i\frac{{\bf p}_{{\rm ph}}^{2}}{2m}t_{0}/\hbar }.
\label{initcond}
\end{align}
Here the wave functions 
\begin{equation}
\Psi _{+}({\bf x},t_{0})=A_{+}\Psi ({\bf x}),
\end{equation}
and 
\begin{equation}
\Psi _{-}({\bf x},t_{0})=A_{-}\Psi ({\bf x}),
\end{equation}
are multiples of the stationary wave function of the parent condensate $\Psi
({\bf x})$. The latter is a solution of the time independent Gross
Pitaevskii equation containing the trapping potential. The amplitudes $A_{+}$
and $A_{-}$ are normalized as $|A_{+}|^{2}+|A_{-}|^{2}=1$.
This ensures that $\int d{\bf x}|\Psi ({\bf x},t_{0})|^{2}$ equals the
number of atoms in the parent condensate, as long as the overlap in momentum
space of the two components on the right hand side of Eq.~(\ref{initcond})
is negligible. 
%The higher order cumulants are assumed to be negligible at
%the initial time $t_{0}$.( {\em What is this comment for?Is it about the
%norm outside the condensate?})

The following analysis of the scattering process is based on our
non-Markovian nonlinear Schr\"{o}dinger equation given by Eq.~(\ref
{NLSnonMarkovEnergBasis}). In position space this equation has the following
form: 
\begin{align}
& i\hbar \frac{\partial }{\partial t}\Psi ({\bf x},t)=H_{{\rm 1B}}({\bf x}%
)\Psi ({\bf x},t)  \nonumber \\
& +\int d{\bf y}_{1}d{\bf y}_{2}d{\bf y}_{3}\int_{t_{0}}^{\infty
}dt_{1}\langle {\bf x},{\bf y}_{3}|T_{{\rm 2B}}^{(+)}(t,t_{1})|{\bf y}_{1},%
{\bf y}_{2}\rangle   \nonumber \\
& \quad \times \Psi ({\bf y}_{1},t_{1})\Psi ({\bf y}_{2},t_{1})\Psi ^{\ast }(%
{\bf y}_{3},t).  \label{NLSnonMarkovPosBasis}
\end{align}
The initial condition Eq.~(\ref{initcond}) implies that the representation 
\begin{align}
\Psi ({\bf x},t)=& \left[ \Psi _{+}({\bf x},t)e^{i{\bf p}_{{\rm ph}}\cdot 
{\bf x}/\hbar }+\Psi _{-}({\bf x},t)e^{-i{\bf p}_{{\rm ph}}\cdot {\bf x}%
/\hbar }\right]   \nonumber \\
& \times e^{-i\frac{{\bf p}_{{\rm ph}}^{2}}{2m}t/\hbar }  \label{IntPic}
\end{align}
provides a suitable interaction picture, where $\Psi _{+}({\bf x},t)$ and $%
\Psi _{-}({\bf x},t)$ are assumed to be slowly varying in space and time.
This interaction picture representation was proposed by Band et al.~in Ref.~
\cite{Band} and is equivalent to the slowly varying envelope
approximation SVEA of nonlinear optics 
\cite{Taylor}. 
Equation (\ref{IntPic}) inserted into Eq.~(\ref
{NLSnonMarkovPosBasis}) yields the following equation of motion: 
\begin{align}
& \left\{ e^{i{\bf p}_{{\rm ph}}\cdot {\bf x}/\hbar }\left[ i\hbar \left( 
\frac{\partial }{\partial t}+\frac{{\bf p}_{{\rm ph}}}{m}\cdot
\bigtriangledown _{{\bf x}}\right) +\frac{\hbar ^{2}}{2m}\Delta _{{\bf x}}%
\right] \Psi _{+}({\bf x},t)\right\}   \nonumber \\
& +\{{\bf p}_{{\rm ph}}\leftrightarrow -{\bf p}_{{\rm ph}}\}  \nonumber \\
& =\int d{\bf y}_{1}d{\bf y}_{2}d{\bf y}_{3}\int_{t_{0}}^{\infty }dt_{1}e^{i%
\frac{{\bf p}_{{\rm ph}}^{2}}{m}(t-t_{1})/\hbar }  \nonumber \\
& \quad \times \langle {\bf x},{\bf y}_{3}|T_{{\rm 2B}}^{(+)}(t,t_{1})|{\bf y%
}_{1},{\bf y}_{2}\rangle   \nonumber \\
& \quad \times \left[ e^{i{\bf p}_{{\rm ph}}\cdot {\bf y}_{1}/\hbar }\Psi
_{+}({\bf y}_{1},t_{1})+e^{-i{\bf p}_{{\rm ph}}\cdot {\bf y}_{1}/\hbar }\Psi
_{-}({\bf y}_{1},t_{1})\right]   \nonumber \\
& \quad \times \left[ e^{i{\bf p}_{{\rm ph}}\cdot {\bf y}_{2}/\hbar }\Psi
_{+}({\bf y}_{2},t_{1})+e^{-i{\bf p}_{{\rm ph}}\cdot {\bf y}_{2}/\hbar }\Psi
_{-}({\bf y}_{2},t_{1})\right]   \nonumber \\
& \quad \times \left[ e^{-i{\bf p}_{{\rm ph}}\cdot {\bf y}_{3}/\hbar }\Psi
_{+}^{\ast }({\bf y}_{3},t)+e^{i{\bf p}_{{\rm ph}}\cdot {\bf y}_{3}/\hbar
}\Psi _{-}^{\ast }({\bf y}_{3},t)\right].  \label{ScattI}
\end{align}
Here, $T_{{\rm 2B}}^{(+)}(t,t_{1})$ is the time dependent transition
operator Eq.~(\ref{timedepTmatrix}) in free space and $\{{\bf p}_{{\rm ph}%
}\leftrightarrow -{\bf p}_{{\rm ph}}\}$ means the exchange of ${\bf p}_{{\rm %
ph}}$ and $-{\bf p}_{{\rm ph}}$ and of $\Psi _{+}$ and $\Psi _{-}$ inside
the bracket. The interaction picture wave functions $\Psi _{\pm }({\bf y}%
_{1},t_{1})$ and $\Psi _{\pm }({\bf y}_{2},t_{1})$ are slowly varying on the
time scale of the two body collisional duration 
and can be evaluated at $t_{1}=t$
on the right hand side of Eq.~(\ref{ScattI}). This leads to the Markov
approximation. In virtue of Eq.~(\ref{defTmatrix}) the remaining time
integral yields 
\begin{equation}
\int_{t_{0}}^{\infty }dt_{1}e^{i\frac{{\bf p}_{{\rm ph}}^{2}}{m}%
(t-t_{1})/\hbar }T_{{\rm 2B}}^{(+)}(t,t_{1})=T_{{\rm 2B}}({\bf p}_{{\rm ph}%
}^{2}/m+i0)  \label{MarkovScatt}
\end{equation}
as soon as $t-t_{0}$ exceeds the two body collisional duration.

To perform the spatial integrations on the right hand side of Eq.~(\ref
{ScattI}) the position representation of the $T$ matrix in Eq.~(\ref
{MarkovScatt}) is transformed to the momentum space and becomes 
\begin{align}
& \langle {\bf x},{\bf y}_{3}|T_{{\rm 2B}}({\bf p}_{{\rm ph}}^{2}/m+i0)|{\bf %
y}_{1},{\bf y}_{2}\rangle   \nonumber \\
& =\frac{1}{(2\pi \hbar )^{6}}\int d{\bf p}_{1}d{\bf p}_{2}d{\bf p}_{3}d{\bf %
p}_{4}\delta ({\bf p}_{1}+{\bf p}_{2}-{\bf p}_{3}-{\bf p}_{4})  \nonumber \\
& \times \left\langle \frac{{\bf p}_{3}-{\bf p}_{4}}{2}\right| \hat{T}_{{\rm %
2B}}\left( \frac{{\bf p}_{{\rm ph}}^{2}}{m}-\frac{({\bf p}_{1}+{\bf p}%
_{2})^{2}}{4m}+i0\right) \left| \frac{{\bf p}_{1}-{\bf p}_{2}}{2}%
\right\rangle   \nonumber \\
& \times e^{i({\bf p}_{4}\cdot {\bf x}+{\bf p}_{3}\cdot {\bf y}_{3})/\hbar
}e^{-i({\bf p}_{1}\cdot {\bf y}_{1}+{\bf p}_{2}\cdot {\bf y}_{2})/\hbar },
\label{momspacerelTmatrix}
\end{align}
where $\hat{T}_{{\rm 2B}}$ is the $T$ matrix which contains only the
relative motion of pairs of atoms. The momentum conserving $\delta $
function stems from the translational invariance of $T_{{\rm 2B}}$ in the
center of mass coordinates. For the experimentally relevant photon momenta $%
{\bf p}_{{\rm ph}}$, the $T$ matrix on the right hand side of Eq.~(\ref
{momspacerelTmatrix}) depends significantly on the related kinetic energy $%
{\bf p}_{{\rm ph}}^{2}/m$ and the contact potential approximation 
Eq.~(\ref{zerorangeappr}) cannot be used. 
In the Markov approximation Eq.~(\ref{ScattI}) then reads 
\begin{align}
& \left\{ e^{i{\bf p}_{{\rm ph}}\cdot {\bf x}/\hbar }\left[ i\hbar \left( 
\frac{\partial }{\partial t}+\frac{{\bf p}_{{\rm ph}}}{m}\cdot
\bigtriangledown _{{\bf x}}\right) +\frac{\hbar ^{2}}{2m}\Delta _{{\bf x}}%
\right] \Psi _{+}({\bf x},t)\right\}   \nonumber \\
& +\{{\bf p}_{{\rm ph}}\leftrightarrow -{\bf p}_{{\rm ph}}\}  \nonumber \\
& =\frac{1}{\sqrt{2\pi \hbar }^{3}}\int d{\bf p}_{1}d{\bf p}_{2}d{\bf p}_{3}d%
{\bf p}_{4}\delta ({\bf p}_{1}+{\bf p}_{2}-{\bf p}_{3}-{\bf p}_{4}) 
\nonumber \\
& \times \left\langle \frac{{\bf p}_{3}-{\bf p}_{4}}{2}\right| \hat{T}_{{\rm %
2B}}\left( \frac{{\bf p}_{{\rm ph}}^{2}}{m}-\frac{({\bf p}_{1}+{\bf p}%
_{2})^{2}}{4m}+i0\right) \left| \frac{{\bf p}_{1}-{\bf p}_{2}}{2}%
\right\rangle   \nonumber \\
& \times \bigg\{\big[\Psi _{+}({\bf p}_{1}-{\bf p}_{{\rm ph}},t)\Psi _{+}(%
{\bf p}_{2}-{\bf p}_{{\rm ph}},t)\Psi _{+}^{\ast }({\bf p}_{3}-{\bf p}_{{\rm %
ph}},t)  \nonumber \\
& \quad +\Psi _{+}({\bf p}_{1}-{\bf p}_{{\rm ph}},t)\Psi _{-}({\bf p}_{2}+%
{\bf p}_{{\rm ph}},t)\Psi _{+}^{\ast }({\bf p}_{3}-{\bf p}_{{\rm ph}},t) 
\nonumber \\
& \quad +\Psi _{-}({\bf p}_{1}+{\bf p}_{{\rm ph}},t)\Psi _{+}({\bf p}_{2}-%
{\bf p}_{{\rm ph}},t)\Psi _{+}^{\ast }({\bf p}_{3}-{\bf p}_{{\rm ph}},t) 
\nonumber \\
& \quad +\Psi _{+}({\bf p}_{1}-{\bf p}_{{\rm ph}},t)\Psi _{+}({\bf p}_{2}-%
{\bf p}_{{\rm ph}},t)\Psi _{-}^{\ast }({\bf p}_{3}+{\bf p}_{{\rm ph}},t)\big]
\nonumber \\
& \quad +\big[{\bf p}_{{\rm ph}}\leftrightarrow -{\bf p}_{{\rm ph}}\big]%
\bigg\}e^{i{\bf p}_{4}\cdot {\bf x}/\hbar },  \label{ScattII}
\end{align}
where $\Psi _{\pm }({\bf p},t)$ denotes the Fourier transform 
\begin{equation}
\Psi _{\pm }({\bf p},t)=\frac{1}{\sqrt{2\pi \hbar }^{3}}\int d{\bf x}e^{-i%
{\bf p}\cdot {\bf x}/\hbar }\Psi _{\pm }({\bf x},t).
\end{equation}

The wave functions $\Psi _{\pm }({\bf x},t)$ are slowly varying in space,
and so their Fourier transform $\Psi _{\pm }({\bf p},t)$ is sharply peaked
at zero momentum (${\bf p}=0$) with a width determined by the initial
momentum spread of the parent condensate. According to Eq.~(\ref{defTmatrix}%
), the two body $T$ matrix $\hat{T}_{2B}(E+i0)$ is slowly varying on the
scale of the energies corresponding to the inverse two body 
collisional duration.
Its plane wave basis matrix elements $\langle {\bf p}_{{\rm out}}|\hat{T}%
_{2B}|{\bf p}_{{\rm in}}\rangle $ are slowly varying in ${\bf p}_{{\rm in}}$
and ${\bf p}_{{\rm out}}$, on the scale of the momentum associated with the
inverse range of the two body interaction potential $V({\bf r})$. This can
be seen through the relation: 
\begin{align}
\langle {\bf r}_{{\rm out}}|\hat{T}_{{\rm 2B}}|{\bf r}_{{\rm in}}\rangle =&
V({\bf r}_{{\rm in}})\delta ({\bf r}_{{\rm in}}-{\bf r}_{{\rm out}}) 
\nonumber \\
& +V({\bf r}_{{\rm out}})\langle {\bf r}_{{\rm out}}|\hat{G}_{{\rm 2B}}|{\bf %
r}_{{\rm in}}\rangle V({\bf r}_{{\rm in}}),
\end{align}
where $\hat{G}_{{\rm 2B}}$ denotes the two body energy-dependent Green's
function for the relative motion only. As a consequence, the plane wave
matrix element of the $T$ matrix $\hat{T}_{2B}$ on the right hand side of
Eq.~(\ref{ScattII}) can be evaluated at the respective momenta determined by
the products of momentum space condensate wave functions and by the momentum
conserving $\delta $ function. Performing the momentum space integrals then
transforms the condensate wave functions $\Psi _{\pm }$ back to position
space and leads to the following equation of motion: 
\begin{align}
& \left\{ e^{i{\bf p}_{{\rm ph}}\cdot {\bf x}/\hbar }\left[ i\hbar \left( 
\frac{\partial }{\partial t}+\frac{{\bf p}_{{\rm ph}}}{m}\cdot
\bigtriangledown _{{\bf x}}\right) +\frac{\hbar ^{2}}{2m}\Delta _{{\bf x}}%
\right] \Psi _{+}({\bf x},t)\right\}   \nonumber \\
& +\{{\bf p}_{{\rm ph}}\leftrightarrow -{\bf p}_{{\rm ph}}\}  \nonumber \\
& =\bigg\{e^{i{\bf p}_{{\rm ph}}\cdot {\bf x}/\hbar }\frac{4\pi \hbar ^{2}}{m%
}\bigg[a_{0}\left| \Psi _{+}({\bf x},t)\right| ^{2}\Psi _{+}({\bf x},t) 
\nonumber \\
& \quad -\left[ f({\bf k}_{{\rm ph}},{\bf k}_{{\rm ph}})+f({\bf k}_{{\rm ph}%
},-{\bf k}_{{\rm ph}})\right] \left| \Psi _{-}({\bf x},t)\right| ^{2}\Psi
_{+}({\bf x},t)\bigg]  \nonumber \\
& \quad +e^{3i{\bf p}_{{\rm ph}}\cdot {\bf x}/\hbar }\langle -2{\bf p}_{{\rm %
ph}}|\hat{T}_{{\rm 2B}}(i0)|0\rangle \Psi _{-}^{\ast }({\bf x},t)\Psi
_{+}^{2}({\bf x},t)\bigg\}  \nonumber \\
& +\left\{ {\bf p}_{{\rm ph}}\leftrightarrow -{\bf p}_{{\rm ph}}\right\}.
\label{ScattIII}
\end{align}
Here, ${\bf k}_{{\rm ph}}={\bf p}_{{\rm ph}}/\hbar $ denotes the wave vector
associated with the photon momentum ${\bf p}_{{\rm ph}}$ and 
\begin{align}
f({\bf k}_{{\rm out}},{\bf k}_{{\rm in}})=& -2\pi ^{2}m\hbar   \nonumber \\
& \times \langle \hbar {\bf k}_{{\rm out}}|\hat{T}_{{\rm 2B}}(\hbar
^{2}k^{2}/m+i0)|\hbar {\bf k}_{{\rm in}}\rangle   \label{ScattAmp}
\end{align}
is the scattering amplitude, where the wave vectors ${\bf k}_{{\rm in}}$ and 
${\bf k}_{{\rm out}}$ are on the energy shell $|{\bf k}_{{\rm out}}|=|{\bf k}%
_{{\rm in}}|=k$. Equation (\ref{ScattIII}) contains two well separated
momentum components on its left hand side and four on the right hand side.
Keeping only the phase matched terms yields a system of two coupled
nonlinear Schr\"{o}dinger equations. These are given by 
\begin{align}
& \bigg[i\hbar \left( \frac{\partial }{\partial t}+\hbar \frac{{\bf k}_{{\rm %
ph}}}{m}\cdot \bigtriangledown _{{\bf x}}\right) +\frac{\hbar ^{2}}{2m}%
\Delta _{{\bf x}}  \nonumber \\
& \quad -\frac{4\pi \hbar ^{2}}{m}a_{0}\left| \Psi _{+}({\bf x},t)\right|
^{2}\bigg]\Psi _{+}({\bf x},t)  \nonumber \\
& =-\frac{4\pi \hbar ^{2}}{m}\left[ f({\bf k}_{{\rm ph}},{\bf k}_{{\rm ph}%
})+f({\bf k}_{{\rm ph}},-{\bf k}_{{\rm ph}})\right]   \nonumber \\
& \qquad \times \left| \Psi _{-}({\bf x},t)\right| ^{2}\Psi _{+}({\bf x},t)
\label{ScattIV}
\end{align}
for the wave function $\Psi _{+}$ and the analogous relation for $\Psi _{-}$
where ${\bf k}_{{\rm ph}}$ and $-{\bf k}_{{\rm ph}}$ as well as $\Psi _{+}$
and $\Psi _{-}$ are exchanged.

As a result of the Markov approximation the $T$ matrix elements in Eq.~(\ref
{ScattIV}) are evaluated on the two-body energy shell and can be represented
in terms of the scattering amplitude Eq.~(\ref{ScattAmp}). 
The experimentally relevant photon momenta are sufficiently small for the
scattering amplitudes on the right hand side of Eq.~(\ref{ScattIV}) to be in
the isotropic limit. According to effective range theory \cite{Newton} $f(%
{\bf k}_{{\rm out}},{\bf k}_{{\rm in}})$ in Eq.~(\ref{ScattAmp}) can then be
expanded in the form: 
\begin{equation}
f({\bf k}_{{\rm out}},{\bf k}_{{\rm in}})=-a_{0}+ia_{0}(ka_{0})+{\cal O}%
(k^{2}).  \label{effrangef}
\end{equation}
Equation (\ref{effrangef}) inserted into Eq.~(\ref{ScattIV}) recovers the
result of Band et al.~in Ref.~\cite{Band}. In the latter article, the two
body energy conservation in the collisions was assumed for physical reasons
and the imaginary part on the right hand side of Eq.~(\ref{effrangef})
included, on an heuristic basis in the Gross Pitaevskii equation, in the
form of Eq.~(\ref{GPE}). The imaginary correction term includes the loss of
those atoms from the condensates that are elastically scattered out of the
forward direction. As reported in Ref.~\cite{Band}, these losses have been
observed experimentally in four wave mixing experiments with Bose
condensates \cite{Deng} and exhibit an excellent agreement with the
predictions of the appropriate generalization of Eq.~(\ref{ScattIV}).

\section{Conclusions}
\noindent
We have shown in this paper how we may systematically obtain a
theoretical description of the dynamics of dilute Bose gases which are
far from their thermal equilibrium. The underlying cumulant 
approach is based on truncating the infinite hierarchy of quantum 
evolution equations for correlation functions of boson field operators in 
accordance with Wick's theorem of statistical mechanics. We extend this 
approach by giving a general scheme for including multiple scattering in
all dynamic equations. This extended approach describes the 
inter-particle collisions in terms of scattering amplitudes 
rather than bare potentials.

The lowest order approximation 
consists in a non Markovian nonlinear Schr\"odinger equation.
When applied to initial conditions close to the thermal equilibrium at
zero temperature our lowest order approach recovers the time 
dependent Gross Pitaevskii equation. 
Applied to the scattering of two condensates at a sharply defined
relative velocity, which is high in comparison with the internal 
velocity spread in each condensate, we have shown that the non Markovian 
nonlinear Schr\"odinger equation goes beyond the Gross Pitaevskii approach 
and describes the experimentally observed collisional loss of condensate 
atoms. Hence, including multiple scattering not simply justifies
the common contact potential approximation but accounts for important  
new physical phenomena.

The second order basic cumulant approach in the contact potential
approximation was shown to be equivalent to the first order time 
dependent Hartree-Fock Bogoliubov theory derived in Ref.~\cite{HollandPRL}.
A detailed description of the physical significance of the corresponding
extended approach, which includes multiple scattering, goes beyond the
scope of this article and will appear elsewhere \cite{TKKB2}.  

Our first order approach should also be applicable when the 
relative bulk motion is still rapid but its velocity is not sharply 
defined. This situation occurs during the collapse of a Bose condensed
gas with a negative $s$ wave scattering length. 
Very recently, the dynamics of a collapse
and the related losses of condensate atoms have been determined 
experimentally in Ref.~\cite{Donley}.

\acknowledgements{\noindent
We would like to gratefully acknowledge inspiring discussions with
Paul Julienne.
This work was supported by the Alexander von Humboldt Foundation and 
the United Kingdom Engineering and Physical Sciences Research Council 
and the European Union.}


\begin{references}

\bibitem{Deng}  L.~Deng, E.~W.~Hagley, J.~Wen, M.~Trippenbach, Y.~Band,
P.~S.~Julienne, J.~E.~Simsarian, K.~Helmerson, S.~L.~Rolston, and
W.~D.~Phillips, Nature (London) {\bf 398}, 218-220 (1999).

\bibitem{Roberts}
J.~L.~Roberts, N.~R.~Claussen, S.~L.~Cornish, E.~A.~Donley, E.~A.~Cornell 
and C.~E.~Wieman, Phys.~Rev~Lett.~{\bf 86}, 4211 (2001).

\bibitem{Donley}
E.~A.~Donley, N.~R.~Claussen, S.~L.~Cornish, J.~L.~Roberts,
E.~A.~Cornell and C.~E.~Wieman, Nature (London) {\b 412}, 295 (2001). 
\bibitem{GZ5}
C.~W.~Gardiner and P.~Zoller, Phys.~Rev.~A {\bf 61}, 033601 (2000)
and references therein.

\bibitem{HWC}
M.~Holland, J.~Williams, and J.~Cooper, Phys.~Rev.~A {\bf 55}, 
3670 (1997).

\bibitem{WWCH}
R.~Walser, J.~Williams, J.~Cooper, and M.~Holland, Phys.~Rev.~A {\bf 59},
3878 (1999).

\bibitem{Fricke}  J.~Fricke, Ann.~Phys.~(N.Y.) {\bf 252}, 479 (1996).

\bibitem{FetterWalecka}  A.~L.~Fetter, J.~D.~Walecka, {\em Quantum Theory of
Many-Particle Systems}, (McGraw-Hill, New York, 1971).

\bibitem{Timmermans}
E.~Timmermans, P.~Tommasini, R.~C\^ot\'e, M.~Hussein, and A.~Kerman, 
Phys.~Rev.~Lett.~{\bf 83}, 2691 (1999).

\bibitem{HollandPRL}  M.~Holland, J.~Park and R.~Walser, Phys.~Rev.~Lett.~%
{\bf 86}, 1915 (2001).

\bibitem{ProukakisBurnett}  N.~P.~Proukakis and K.~Burnett,
J.~Res.~Natl.~Inst.~Stand. Technol.~{\bf 101}, 457 (1996).

\bibitem{Newton}  R.~G.~Newton, {\em Scattering Theory of Waves and Particles%
} (Springer, New York, 1982).

\bibitem{coherentstate}  The density matrix of a purely condensed ideal gas
is given by a pure coherent state of the lowest trap mode. With the latter
chosen as the initial state of the interacting dilute gas all initial normal
ordered correlation functions factorize into products of the condensate wave
function or its conjugate, and Eq.~(\ref{defcumI}) shows recursively that
all respective cumulants except the condensate wave function vanish.

\bibitem{Cornish}  S.~L.~Cornish, N.~R.~Claussen, J.~L.~Roberts,
E.~A.~Cornell, and C.~E.~Wieman, Phys.~Rev.~Lett.~{\bf 85}, 1795 (2000). 
%\bibitem{Tiesinga}
%E.~Tiesinga, C.~J.~Williams, F.~H.~Mies, and P.~S.~Julienne, 
%Phys.~Rev.~A {\bf 61}, 63416 (2000).

\bibitem{Band}  Y.~B.~Band, M.~Trippenbach, J.~P.~Burke, and P.~S.~Julienne,
Phys.~Rev.~Lett.~{\bf 84}, 5462 (2000).

\bibitem{Taylor}
See, e.g., E.~M.~Dianov, A.~B.~Grudinin, A.~M.~Prokhorov and V.~N.~Serkin in
{\em Optical Solitons -- Theory and Experiment},
edited by J.~R.~Taylor
(Cambridge University Press, Cambridge, 1992).

\bibitem{TKKB2}
T.~K\"ohler and K.~Burnett (unpublished).
\end{references}
\end{document}